# Effect of Data Preprocessing on Software Effort Estimation

Sumeet Kaur Sehra
Assistant Professor
Department of CSE
GNDEC Ludhiana

Jasneet Kaur
Senior Faculty
NIELIT
Chandigarh

Sukhjit Singh Sehra
Assistant Professor
Department of CSE
GNDEC Ludhiana

## ABSTRACT
Software effort estimation requires high accuracy, but accurate estimations are difficult to achieve. Increasingly, data mining is used to improve an organization's software process quality, e.g. the accuracy of effort estimations .There are a large number of different method combination exists for software effort estimation, selecting the most suitable combination becomes the subject of research in this paper. In this study, three simple preprocessors are taken (none, norm, log) and effort is measured using COCOMO model. Then results obtained from different preprocessors are compared and norm preprocessor proves to be more accurate as compared to other preprocessors.

## 1. INTRODUCTION
The effort invested in a software project is probably one of the most important and most analyzed variables in recent years in the process of project management [1] .From the beginning of software engineering as a research area more than three decades ago, several development effort estimation methods and process have been proposed. Being able to choose the most suitable software development effort estimator for the local software projects remains elusive for many project managers [2]. For decades, researchers have been searching for the "best" software development effort estimator [3]. Increasingly, data mining is used to improve an organization's software process quality, e.g. the accuracy of effort estimations. In data mining process data is collected from projects, and data miners are used to discover beneficial knowledge.Datapre-proces is ingisanoftenneglectedbutimportantstepinthedataminingprocess. As rawdataishighly susceptible to noise, missing values, andinconsistency.TheQualityofdataaffectsthedataminingresults .Inordertohelpimprovethequalityofthedataand,consequently,oft heminingresultsrawdataispre-processedsoastoimprovetheEfficiencyandeaseoftheminingproc ess. Also ifthereismuchirrelevantandredundantinformationpresentornois yandunreliabledata,thenknowledgediscoveryduringthetrainingp haseismoredifficult.Datapreparationandfilteringstepscantakeco nsiderableamountofprocessing time.Datapre-processingincludes cleaning, normalization,

Transformation,featureextractionandselection,etc.Theproductof datapre-processingisthefinaltrainingset [4]. In this study we investigate that data preprocessing is one of the
Most critical step in data mining process which deals with preparation and transformation of Initial data set.

## 2. RELATED WORK
In the software Engineering literature there are so many models that are used to estimate the effort. This section provides some information of various software effort estimation models to be used in this work

### 2.1 Cocomo Basic Model
The Constructive Cost Model (COCOMO) is an algorithmic software cost estimation model developed by Barry W. Boehm. The model uses a basic regression formula with parameters that are derived from historical project data and current project characteristics [5]. Basic COCOMO computes software development effort (and cost) as a function of program size. Program size is expressed in estimated thousands of Kilo lines of code (KLOC).COCOMO applies to three classes of software projects [2]:

- Organic projects - "small" teams with "good" experience working with "less than rigid" requirements
- Semi-detached projects - "medium" teams with mixed experience working with a mix of rigid and less than rigid requirements
- Embedded projects - developed within a set of "tight" constraints. It is also combination of organic and semi-detached projects. (hardware, software, operational, ...)

The basic COCOMO equation for calculating effort take the form

Effort Applied (E) = a $(KLOC)^b$ (1)

Where, KLOC is the estimated number of delivered lines (expressed in thousands) of code for project. The coefficients *a*, *b* are given in the following table [2]:

**Table 1. COCOMO Model**

| Software project | a | b |
|---|---|---|
| Organic | 2.4 | 1.05 |
| Semi-detached | 3.0 | 1.12 |
| Embedded | 3.6 | 1.21 |

# 3. EVALUATION CRITERIA

For the purpose of validating and evaluating the new methodology, the basic necessity is to measure how accurate the estimations are. There are various approaches used by researchers to measure the accuracy of effort. The Basic COCOMO software effort model with development mode organic is taken for calculating the effort by using equation

$$Effort = 2.4(KLOC)^{1.05} \quad (2)$$

## 3.1. Data preprocessing

Data pre-processing is an important step in the data mining process. The phrase "garbage in, garbage out" is particularly applicable to data mining and machine learning projects. Data-gathering methods are often loosely controlled, resulting in out-of-range values (e.g., Income: −100), impossible data combinations (e.g., Gender: Male, Pregnant: Yes), missing values, etc. Analyzing data that has not been carefully screened for such problems can produce misleading results. Thus, the representation and quality of data is first and foremost before running an analysis. Data preprocessing includes:-

- Cleaning
- Normalization
- Transformation
- Feature extraction and selection, etc.

In this study, we investigate: – Three preprocessors [3]:

a) **None Technique: -**None is the simplest preprocessor- all values are unchanged. In this the effort is directly calculated using Cocomo model equation (2).

b) **Norm Technique: -** With the norm preprocessor (max-min normalization), numeric values are normalized to a 0-1 interval using formula:

$$\begin{aligned}&\text{NormalizedValue}\\&= \frac{\text{Actual Value} - \min(\text{All Values})}{\text{Max(All Values)} - \text{Min(All Values)}}\end{aligned} \quad (3)$$

Where

- Actual value: - current valued to be mapped
- Min: - minimum value from particular column to be normalized in given data set
- Max: - Max value from particular column to be normalized in given data set

In Norm techniques first the Dataset is Normalized using equation (3) and then effort is calculated.

c) **Log Technique: -** With the log preprocessor, all numeric are replaced with their logarithm. This logging Procedure Minimizes the effects of the occasional very large numeric value.

Log value = $\log_2$(Actual value)(4)

# 4. DATA COLLECTION

The data is collected of IVR application[2] though survey from BPO / software Industry where IVR application is developed with questionnaires directly to the company's project managers and senior software development professionals. Researcher also arranged interview sessions over telephone with surveyed company's personnel to know the actual process capability of the company. Researcher asked the set of questions during the phone interview as well as email session's .Question sets are related to IVR software application An IVR system (IVRS) accepts a combination of voice telephone input and touch-tone keypad selection and provides appropriate responses in the form of voice, fax, call back or other any media. An IVR system consists of telephony equipment, software applications, and a database. Due to Company security and policy the author could not show the name of project but it is indicated as Project No. Actual Size, Actual Effort is measured in line of code metrics, person-month and month respectively as shown in table below. Effort is the number of labor units required to complete an activity.

**Table 2:  Real data of IVR projects**

| Project No | Actual Size(KLOC) | Actual Effort |
|---|---|---|
| 1 | 16.2 | 86.1 |
| 2 | 5.34 | 24.02 |
| 3 | 7.6 | 36.05 |
| 4 | 4.7 | 20.74 |
| 5 | 3.1 | 12.85 |
| 6 | 5.2 | 23.3 |
| 7 | 6.8 | 31.72 |
| 8 | 6.4 | 29.59 |
| 9 | 7.2 | 33.88 |
| 10 | 5.4 | 24.34 |
| 11 | 8.5 | 41.01 |
| 12 | 7.8 | 37.15 |
| 13 | 12.5 | 63.9 |
| 14 | 10.4 | 51.71 |
| 15 | 9.5 | 46.6 |
| 16 | 3.4 | 14.29 |
| 17 | 6.8 | 31.73 |
| 18 | 5.8 | 26.42 |
| 19 | 7.4 | 34.96 |
| 20 | 7.2 | 33.88 |
| 21 | 8.6 | 41.56 |

| 22 | 6.4 | 29.59 |
| 23 | 10.6 | 52.86 |
| 24 | 6.3 | 29.06 |

| 28.6274 | 1.3127 | 5.9147 |
| 16.5775 | 0.6262 | 4.5541 |

## 5. EXPIREMENTAL RESULTS

The data preprocessing techniques are applied on actual data set. In case of norm and log techniquesthe KLOC is calculated using equations3 and 4 where as in case of none data preprocessing technique KLOC value is not affected.Thus the corresponding effort is calculated by using equation 2 as shown below

**Table 3: Effort estimated after Data preprocessing**

| Estimated Effort (Using None Technique) | Estimated Effort (Using Norm Technique) | Estimated Effort (Using Log Technique) |
|---|---|---|
| 44.6891 | 2.2334 | 7.0353 |
| 13.9357 | 0.4768 | 4.1256 |
| 20.1867 | 0.8311 | 5.0427 |
| 12.1875 | 0.3784 | 3.7961 |
| 7.873 | 0.1388 | 2.7322 |
| 13.5524 | 0.4552 | 4.0569 |
| 17.9616 | 0.7047 | 4.7528 |
| 16.8539 | 0.6418 | 4.5951 |
| 19.0726 | 0.7678 | 4.9017 |
| 14.1002 | 0.4861 | 4.1545 |
| 22.7039 | 0.9745 | 5.3353 |
| 20.7449 | 0.8629 | 5.1106 |
| 34.0382 | 1.6224 | 6.3492 |
| 28.0606 | 1.2803 | 5.8646 |
| 25.5165 | 1.135 | 5.6268 |
| 8.6749 | 0.1828 | 2.9669 |
| 17.9616 | 0.7047 | 4.7528 |
| 15.1988 | 0.5481 | 4.3395 |
| 19.6293 | 0.7994 | 4.9731 |
| 19.0726 | 0.7678 | 4.9017 |
| 22.9845 | 0.9905 | 5.3659 |
| 16.8539 | 0.6418 | 4.5951 |

## 5.1 Graphical Representation

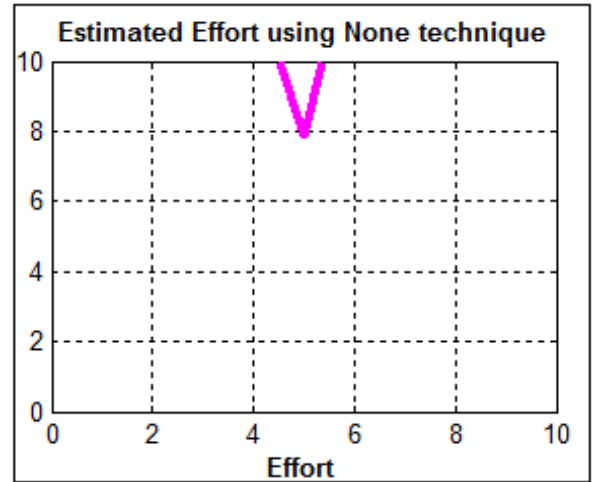

Also graphically the results obtained by applying data preprocessing techniques (none technique, norm technique and log technique) are shown below:-

**None Technique**: - As shown below is effort when calculated

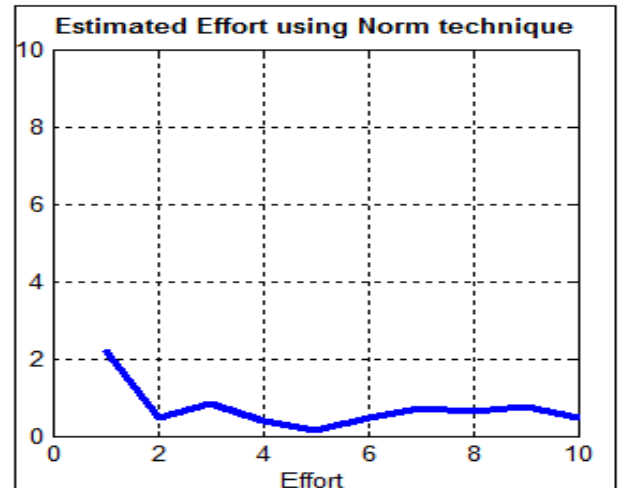

using none technique

**Figure 1: Estimated effort using None Technique**

**Norm Technique**: - As shown below is effort when calculated using norm technique.

**Figure 2: Estimated effort using Norm Technique**

**Log Technique**: - As shown below is effort when calculated using log technique

**Figure 3: Estimated effort using Log Technique**

## 6. CONCLUSIONS

Producing accurate software estimates has always been a challenge, where no one method has established itself to the fullest to consistently deliver an accurate estimate. Since effort is a continuous attribute, typically some error is expected. However, if estimate is far from actual value, e.g,more than 25 %,then estimate cannot be considered "accurate" [6]. The evaluations have revealed that results obtained by applying normdata preprocessing techniques are more accurate (less than 25 %).

Also graphical reprenstation shows that the estimated effort using norm techniques gives more stable values as compared to estimated effort using Log technique and None technique. Thus Norm technique is chosen as best data preprocessing technique for effort estimation.

This methodology can further be explored on some other large datasets with in order to further enhance the validity of the produced results.

Management Systems (IJDMS), Vol.3, No.1, February

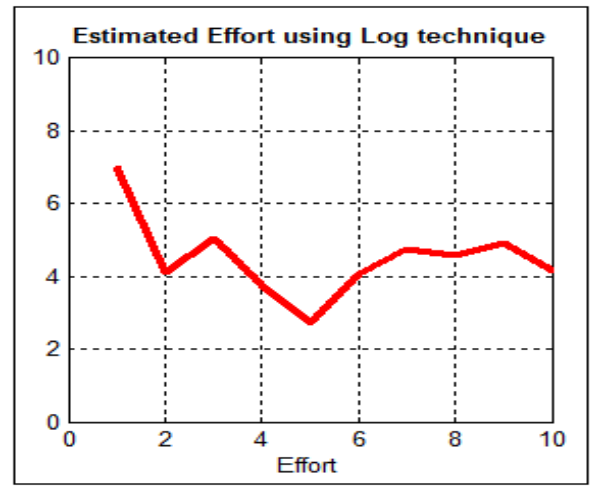

2011.

## 7. REFRENCES